\documentclass[pre,twocolumn,aps,showpacs]{revtex4}
\newcommand{\bec}[1]{\mbox{\boldmath $ #1$}}
\usepackage{graphicx}
\begin{document}
\title{Mean-field dynamos in random Arnold-Beltrami-Childress and
Roberts flows}
\author{Nathan Kleeorin}
\email{nat@bgu.ac.il}
\author{Igor Rogachevskii}
\email{gary@bgu.ac.il}
\homepage{http://www.bgu.ac.il/~gary}
\affiliation{Department of Mechanical Engineering,
 The Ben-Gurion University of the Negev,
 POB 653, Beer-Sheva 84105, Israel}
\author{Dmitry Sokoloff}
\email{sokoloff@dds.srcc.msu.su}
\affiliation{Department of Physics, Moscow State University,
Moscow 119992, Russia}
\author{Dmitry Tomin}
\email{dtomin@gmail.com}
\affiliation{Department of Mechanics and Mathematics, Moscow State University, Moscow 119992, Russia}
\date{\today}
\begin{abstract}
We study magnetic field evolution in flows with fluctuating in time governing parameters in electrically conducting fluid. We use a standard mean-field approach to derive equations for large-scale magnetic field for the fluctuating $ABC$-flow as well as for the fluctuating Roberts flow. The derived mean-field dynamo equations have growing solutions with growth rate of the large-scale magnetic field which is not controlled by molecular magnetic diffusivity. Our study confirms the Zeldovich idea that the nonstationarity of the fluid flow may remove the obstacle in large-scale dynamo action of classic stationary flows.
\end{abstract}


\maketitle

\section{Introduction}

Many celestial bodies including the Earth, the Sun and the Milky Way
have magnetic fields with spatial scales which are
much larger then the basic (maximum) scale of turbulence or turbulent
convection. It has been widely recognized that these magnetic fields
originate due to the mean-field dynamo based on a joint action of
differential rotation and $\alpha$-effect operating in the
mirror-asymmetric turbulence or turbulent convection (see, e.g.,
\cite{M78,P79,KR80,ZRS83,RSS88,RH04,O03,BS05}). More
recent models of large-scale dynamos include also such phenomena as
shear-current effect, kinetic helicity fluctuations, etc.
(see, e.g., \cite{RK03,VB97,SOK97,SIL00,PROC07,KR08}). Corresponding
dynamo models for particular celestial bodies result
in magnetic fields configurations which are compatible with available
phenomenology at least in a crude approximation (see, e.g., \cite{BECK96}).

On the other hand, many important features in astrophysical dynamos
remain unclear (see, e.g., \cite{K99}). In particular, many
realistic dynamo models contain a concept of turbulence or turbulent
convection which is considered as a flow random in space and time.
This concept that originated in Kolmogorov theory (see, e.g.,
\cite{MY75,Mc90,frish}), is accepted by many experts. However, we
have to appreciate the fact that from many points of view it is at
least helpful to reserve a possibility to present astrophysical
flows as deterministic solutions of the Navier-Stokes equation.

Arnold suggested to mimic the magnetic field generation in a random
flow by a dynamo action of a special flow known today as $ABC$
(Arnold-Beltrami-Childress) flow (see, e.g.,
\cite{CG95,ZRS90,A65,H66,HC70}). The point is that a conventional
mean-field dynamo requires a nonzero kinetic helicity, i.e. a
correlation between velocity field $\bf v$ and ${\rm curl} \, {\bf
v}$. On the other hand, there are deterministic flows  (Beltrami
fields) whereby $\bf v$ is parallel to ${\rm curl} \, {\bf v}$. If
we accept the concept of the $\alpha$-effect and the kinetic
helicity for deterministic flows, their kinetic helicity is maximum
so these flows can be suitable for a generation of a large-scale
magnetic field. Simultaneously, Arnold stressed that the Beltrami
velocity fields provide an effective magnetic lines stretching. The $ABC$-flow is a simple example of Beltrami flows. Speaking in modern terms, the $ABC$-flow provides a simple example of dynamical chaos.

The problem however is that this flow is enough complicated, so that
the induction equation cannot be generally solved analytically.
Arnold with coauthors  suggested an artificial example of chaotic
flow in a specially chosen Riemannian space for which the induction
equation can be solved analytically (see \cite{AZ81}). It was
demonstrated in \cite{AZ81} that a large-scale magnetic field can be
generated by this flow. On the other hand, numerical experiments
(see, e.g., \cite{AK83,GPS92,GF86}) also show that the $ABC$-flow
can excite a magnetic field. However, the generated magnetic field
is quite remote from that which astrophysicists are inclined to refer as a large-scale magnetic field. Its spatial configuration looks as a
combination of cigar-like structures, which spatial scale in the
cross-section is controlled by magnetic Reynolds number.

In spite of various impressive results in solving particular
problems for dynamos in stationary deterministic flows (see, e.g.,
\cite{CG95,ZRS90,GF86,DF86,GF87,PP94,GKR94,GKR95}) no particular
deterministic flow is yet known to mimic a self-excitation of the
large-scale magnetic field which widely occurs in random flow.
This situation looks obviously unsatisfactory.

It was suggested by Zeldovich that a nonstationarity of a fluid flow
is important feature required for the generation of the mean
magnetic field (see, e.g., \cite{ZRS83}). Since the induction
equation for a nonstationary flow is even more complicated for
analytical study, it was quite difficult to provide a convincing
support for this idea. It looks reasonable in this context to
introduce time-dependence in the $ABC$-flow (or other deterministic
velocity field) as a noise added to numerical parameters (here $A$,
$B$ and $C$) which govern the flow. This idea as well as possible
approaches have been known for years, however the bulk of analytical
work required was quite large, so no particular solution has been
suggested until now.

The goal of this study is to revise this old and intriguing problem.
This paper is organized as follows. In Sect.~II the mean-field approach
for a flow with random coefficients is formulated. In Sect.~III the
electromotive force is determined for random $ABC$ and Roberts flows.
Finally, we discuss obtained results and draw conclusions in Sec.~IV.
In Appendixes~A and~B we discuss spatial averaging of the mean-field
equation and present details of calculations of the electromotive
force for the $ABC$-flow as well as for the Roberts flow with random coefficients.

\section{Mean-field approach for a flow with random coefficients}

The problem under consideration requires quite a lot of simple
however bulky algebra so we describe here the strategy of the
research undertaken and present results for particular flows. To
be specific, we describe the strategy using $ABC$-flow
which is determined as
\begin{eqnarray}
v_x &=& B \cos y + C \sin z \;, \quad v_y = C \cos z + A \sin x \;,
\nonumber\\
v_z &=& A \cos x + B \sin y \;, \label{B1}
\end{eqnarray}
where $x, y, z$ are measured in units of the characteristic scale $l$ of the velocity field variations. We consider the flow (\ref{B1}) as given and study a kinematic dynamo problem. We do not discuss here how the flow can be excited in practice or what is a nonlinear stage of
dynamo action. We highly appreciate the importance of these
questions (see, e.g., \cite{BS05} and references therein). From our point of view a straightforward way to resolve these
questions is to return to the concept of turbulence as a field that is
random in space and time. We appreciate as well that in practice a kinetic helicity is far to be maximum possible (i.e. realistic
velocity fields have kinetic helicity that is much lower then that for
the Beltrami field).

A first approach applied here is to consider the flow determined by Eq.~(\ref{B1}) with random coefficients $A,B,C$ as a particular example of inhomogeneous and anisotropic random velocity field, and
to apply corresponding expressions for electromotive force $\bec{\cal E}$ known in the literature (see, e.g., \cite{R80,RK97,RKR03}). This approach is practical in the sense that it requires a minimum bulk of algebra and allows a clear identification of various terms which appear in the resulting mean-field equation. This is a reason why we use this approach in the present study. On the other hand, we recognize that the above approach is not fully convincing from the viewpoint of high-brow probability theory. The point is that we accept here that the flow under consideration can mimic a turbulent flow while it is something what we have to proof.

This is a reason why we also need to use another approach consisting
in a direct averaging of the induction equation and not using the
known expressions for electromotive force $\bec{\cal E}$. In the
second approach we have to use a particular model of random $A,B,C$
in form of the $\delta$-correlated in time random processes and
reproduce just from the beginning the procedure of derivation of
mean-field equations for the magnetic field in such flow. The second
approach requires much more algebra and create various problems with
identifications of terms in the resulting equation. However the
final results occur to be identical to those obtained using the
first approach. Comparing the results of these two approaches, one
have to bear in mind that the equations exploited in both cases
suppose different normalization for velocity field (up to the factor
$1/2$ in resulting expressions). We avoid to give here the bulky,
however important calculations used in the second approach and
present them in a separate study \cite{tomin}.

The main result in both approaches is a governing equation for the
magnetic field averaged over the ensemble of realizations of
$A,B,C$. Because the velocity field investigated is far to be
statistically homogeneous and isotropic, the resulting equation is
quire bulky and hardly can be solved analytically. A numerical
approach to this equation looks not easier then that one for initial
equations. We are interesting however in large-scale solutions for
the equations. To this end, we perform an additional spatial
averaging to remove small-scale noise from the solutions and keep
large-scale properties of the solutions only. In practice, we remove
terms like $\sin x$ and $\cos x$ and replace terms like $\sin^2 x$
and $\cos^2 x$ by $1/2$. As a result, we arrive at simple equations
with constant coefficients which we solve in Fourier space.

We perform our analysis for two velocity fields which look
instructive for our problem. One is the above discussed $ABC$-flow
where fluctuations of governing parameters are implied.
We choose as a counterpart another well-known flow suggested for dynamo problem by Roberts \cite{RO70}:

\begin{eqnarray}
v_x &=& - C \sin x \cos y \;, \quad v_y = C \cos x \sin y \;,
\nonumber\\
v_z &=& C \sin x \sin y \;, \label{C1}
\end{eqnarray}
and consider the governing parameter $C$ as random.

The $ABC$ flow and the Roberts flow demonstrate properties which are to some extent opposite from the viewpoint of topological fluid dynamics (see, e.g., \cite{AR-1,AK-2}). In particular, the $ABC$-flow looks as a most advantageous velocity field in sense that the large-scale field generation becomes independent on details of the flow geometry.
The $ABC$-flow is a Beltrami flow, i.e., vectors $\bf v$ and ${\rm curl}\, {\bf v}$ are parallel. Such velocity field provide exponential stretching of magnetic lines.

In contrast, Roberts flow is not a Beltrami flow, i.e.,
two-component vectors $({\rm curl}_x {\bf v}, {\rm curl}_y {\bf v})$
and $(v_x, v_y)$ are parallel, however ${\rm curl}_z {\bf v} =0$
while $v_z \ne 0$. It means that the stationary Roberts flow (i.e., constant in time governing parameter $C$) do not
provide exponential stretching of magnetic lines. On the other hand, we demonstrate in this study that the Roberts flow becomes a dynamo when the coefficient $C$ fluctuates in time (see Sect.~III).

In the next section we investigate how fluctuations of the governing parameters in these velocity fields affect the large-scale dynamo action.

\section{Electromotive force in anisotropic velocity field}

The electromotive force in anisotropic velocity field is determined by the following equation:
\begin{eqnarray}
{\cal E}_{i} = {\alpha}_{ij} b_{j} - {\eta}_{ij} (\bec{\nabla} {\bf \times} {\bf b})_{j} + ({\bf V}^{\rm eff} {\bf \times} {\bf b})_{i}
- {\kappa}_{ijk} ({\partial b})_{jk} \;,
\nonumber\\
\label{R1}
\end{eqnarray}
where ${\bf b} = \langle {\bf H} \rangle$ (in the context of the $ABC$-flow the angular brackets denote averaging over random coefficients $A, B, C$), and
\begin{eqnarray}
\alpha_{ij} &=& - {\tau \over 2} \big(\varepsilon_{imn} \langle  v_n {\bf  \nabla}_{j} v_m \rangle + \varepsilon_{jmn}
\langle v_n {\bf  \nabla}_{i} v_m \rangle\big) \;,
\label{R2} \\
\eta_{ij} &=& {\tau \over 2} \big(\langle {\bf v}^2 \rangle \delta_{ij} - \langle v_i  v_j \rangle\big) \;,
\label{R3} \\
V_i^{\rm eff} &=& - {1 \over 2} {\bf \nabla}_{m} \tau \langle v_m v_i \rangle \;,
\label{R4} \\
{\kappa}_{ijk} &=& - {\tau \over 2} \big(\varepsilon_{ijm} \langle v_m  v_k \rangle + \varepsilon_{ikm} \langle v_m  v_j \rangle\big) \;,
\label{R5}
\end{eqnarray}
where $\tau$ is the characteristic time-scale of the random velocity field, $\delta_{ij}$ is the Kronecker tensor, and $\varepsilon_{ijk}$ is the fully antisymmetric Levi-Civita tensor. Equations~(\ref{R1})-(\ref{R5}) have been derived by the following procedures: (i) the second order correlation approximation (or first-order smoothing) in \cite{R80} that is valid for small Reynolds numbers; (ii) the path integral approach in \cite{RK97} for the delta-correlated in time random velocity field; (iii) the tau-approach in \cite{RKR03} that is valid for large Reynolds numbers.

The spatial averaging $\langle ... \rangle^{(sp)}$ of the induction
equation for the field ${\bf b}$ yields equation for the mean
magnetic field ${\bf B} = \langle {\bf b} \rangle^{(sp)}$. In
spatial averaging the effects, $\langle \tilde {\alpha}_{ij} \tilde
b_{j} \rangle^{(sp)}$, $\,\langle \tilde {\eta}_{ij} (\bec{\nabla}
{\bf \times} \tilde {\bf b})_{j} \rangle^{(sp)}$, $\,\langle \tilde
{\bf V}^{\rm eff} {\bf \times} \tilde {\bf b} \rangle^{(sp)}$ and
$\langle \tilde {\kappa}_{ijk} ({\partial \tilde b})_{jk}
\rangle^{(sp)}$, should be also taken into account. Here
${\eta}_{ij} = \langle {\eta}_{ij} \rangle^{(sp)} + \tilde
{\eta}_{ij}$ and $\langle \tilde {\eta}_{ij} \rangle^{(sp)}=0$, and
similarly for other tensors (tilde denotes spatially variable part
of the tensor). The magnetic field $\tilde {\bf b}$ is determined by
the following equation:

\begin{eqnarray}
&& \varepsilon_{mni} \nabla_n  \big[\langle {\alpha}_{ij} \rangle^{(sp)} \tilde b_{j} -
\langle {\eta}_{ij} \rangle^{(sp)} (\bec{\nabla} {\bf \times} \tilde {\bf b})_{j}
\nonumber\\
&& \quad \quad + (\langle {\bf V}^{\rm eff} \rangle^{(sp)} {\bf \times} \tilde {\bf b})_{i} -
\langle {\kappa}_{ijk} \rangle^{(sp)} ({\partial \tilde b})_{jk}\big] = I_m \;,
\label{R6}\\
&& I_m = - \varepsilon_{mni} \nabla_n \big[\tilde{\alpha}_{ij} B_{j} - \tilde{\eta}_{ij} (\bec{\nabla}
{\bf \times} {\bf B})_{j} + (\tilde{\bf V}^{\rm eff} {\bf \times} {\bf B})_{i}
\nonumber\\
&& - \tilde{\kappa}_{ijk} ({\partial B})_{jk} \big] \; .
\label{R7}
\end{eqnarray}
The solution of Eq.~(\ref{R6}) allows us to determine the second moments $\langle \tilde {\alpha}_{ij} \tilde b_{j} \rangle^{(sp)}$, $\, \langle \tilde {\eta}_{ij} (\bec{\nabla} {\bf \times} \tilde {\bf b})_{j} \rangle^{(sp)}$ and $\langle \tilde {\kappa}_{ijk} ({\partial \tilde b})_{jk} \rangle^{(sp)}$ and to derive equation for the mean magnetic field ${\bf B}$.

\subsection{The $ABC$-flow with random coefficients}

Let us consider the $ABC$-flow that is determined by Eq.~(\ref{B1}), where the coefficients $A, B, C$ are the random variables. Using Eq.~(\ref{R2}) we determine the tensor $\alpha_{ij}$ for the $ABC$-flow:
\begin{eqnarray}
\alpha_{xx} &=& - {\tau \sigma_{_{A}} \over l}  \,, \; \;  \alpha_{yy} = - {\tau \sigma_{_{B}} \over l} \,,
\; \; \alpha_{zz} = - {\tau \sigma_{_{C}} \over l} \,,
\label{B2}
\end{eqnarray}
and $\alpha_{ij} = 0$ for $i\not=j$, where $\sigma_{_{A}} = \langle A^2 \rangle$, $\, \langle A \rangle = 0$  and similarly for random variables $B, C$. The tensor $\alpha_{ij}$ determines the generation of magnetic field. For the $ABC$-flow with random coefficients $\tilde {\alpha}_{ij} =0$ and the effective drift velocity of the magnetic field ${\bf V}^{\rm eff} = 0$ [see Eqs.~(\ref{R4}) and~(\ref{B3}) in Appendix A].
Note that the condition $\langle A \rangle = 0$ implies that
we consider a fluctuating $ABC$-flow with vanishing mean
velocity. In other words, we choose parametrization of the random
flow in such a way to stress the difference between the flow under
consideration and the conventional stationary $ABC$-flow.

Using Eqs.~(\ref{R3}) and~(\ref{B3}) we get the symmetric tensor $\eta_{ij}$ for the $ABC$-flow:
\begin{eqnarray}
\eta_{xx} &=& {\tau \over 2} \, \big[\sigma_{_{A}} + \sigma_{_{B}} \, \sin^2 y + \sigma_{_{C}} \, \cos^2 z\big] \;,
\nonumber\\
\eta_{yy} &=& {\tau \over 2} \, \big[\sigma_{_{B}} + \sigma_{_{C}} \, \sin^2 z + \sigma_{_{A}} \, \cos^2 x \big] \;,
\nonumber\\
\eta_{zz} &=& {\tau \over 2} \, \big[\sigma_{_{C}} + \sigma_{_{A}} \, \sin^2 x + \sigma_{_{B}} \, \cos^2 y\big] \;,
\nonumber\\
\eta_{xy} &=& - {\tau\sigma_{_{C}} \over 4} \, \sin (2 z)
\;,
\quad
\eta_{xz} = - {\tau\sigma_{_{B}} \over 4} \, \sin (2 y) \;,
\nonumber\\
\eta_{yz} &=& - {\tau\sigma_{_{A}} \over 4} \, \sin (2 x) \; .
\label{B4}
\end{eqnarray}
Equation~(\ref{R5}) allows us to determine the tensor ${\kappa}_{ijk} ({\partial b})_{jk}$:
\begin{eqnarray}
{\kappa}_{xjk} ({\partial b})_{jk} &=& {\tau \over 2} \big[u_{yk} ({\partial b})_{zk} - u_{zk} ({\partial b})_{yk} \big] \;,
\nonumber\\
{\kappa}_{yjk} ({\partial b})_{jk} &=& {\tau \over 2} \big[u_{zk} ({\partial b})_{xk} - u_{xk} ({\partial b})_{zk} \big] \;,
\nonumber\\
{\kappa}_{zjk} ({\partial b})_{jk} &=& {\tau \over 2} \big[u_{xk} ({\partial b})_{yk} - u_{yk} ({\partial b})_{xk} \big] \;,
\label{B5}
\end{eqnarray}
where the velocity fluctuations, $u_{ij} = \langle v_i v_j \rangle$, are determined by Eq.~(\ref{B3}). The tensors $\eta_{ij}$ and ${\kappa}_{ijk}$ contribute to the magnetic eddy diffusivity caused by random motions of the conducting fluid. This effect limits the dynamo growth rate of magnetic field.

The spatial averaging $\langle ... \rangle^{(sp)}$ of the electromotive force yields the following equation (see Appendix A):
\begin{eqnarray}
\langle \bec{\cal E} \rangle^{(sp)} = - \tau \sigma_{_{A}} \big({\bf B} \, l^{-1} + \bec{\nabla} {\bf \times} {\bf B} \big) \;,
\label{B6}
\end{eqnarray}
where we considered the case $\sigma_{_{A}} = \sigma_{_{B}} = \sigma_{_{C}}$. Note that there are generally two contributions to the electromotive force caused by the spatial averaging $\langle ... \rangle^{(sp)}$. Indeed, the direct contribution to the electromotive force, $\bec{\cal E} ^{(1)}$, is due to the spatial averaging of Eqs.~(\ref{B2})-(\ref{B5}), while the second contribution to the electromotive force, $\bec{\cal E} ^{(2)}$, is caused by the second moments $\langle \tilde {\alpha}_{ij} \tilde b_{j} \rangle^{(sp)}$, $\, \langle \tilde {\eta}_{ij} (\bec{\nabla} {\bf \times} \tilde {\bf b})_{j} \rangle^{(sp)}$, etc. However, for the $ABC$-flow $\bec{\cal E} ^{(2)}$ vanishes (see Appendix A), while for the Roberts flow this contribution is not zero. Therefore, the equation for the mean magnetic field reads:
\begin{eqnarray}
{\partial {\bf B} \over \partial t} = - {\tau \sigma_{_{A}} \over l} \big( \bec{\nabla} \times {\bf B} \big) +
\big(\eta + \tau \sigma_{_{A}}\big) \, \Delta {\bf B}  \;,
\label{B7}
\end{eqnarray}
where $\eta$ is the magnetic diffusion due to the electrical conductivity of the fluid.

\subsection{The random Roberts flow}

Now let us consider the random Roberts flow~(\ref{C1}), that can be rewritten in the form: ${\bf v} = C \, \big({\bf e} + {\bf e} {\bf \times} \bec{\nabla} \big)\, \sin x \, \sin y$, where ${\bf e}$ is the unit vector in $z$-direction and $C$ is the random variable. Using Eq.~(\ref{R2}) we get the symmetric tensor $\alpha_{ij}$:
\begin{eqnarray}
\alpha_{xx} &=& {\tau \sigma \over l}  \, \sin^2 y \;, \quad \alpha_{yy} = {\tau \sigma \over l} \, \sin^2 x \;,
\nonumber\\
\alpha_{xz} &=& {\tau \sigma \over 4l} \, \sin (2 y)  \;,
\quad
\alpha_{yz} = {\tau \sigma \over 4l} \, \sin (2 x)  \;,
\nonumber\\
\alpha_{xy} &=& \alpha_{zz} = 0 \;,
\label{C2}
\end{eqnarray}
where $\sigma = \langle C^2 \rangle$ and $\langle C \rangle = 0$. We choose parametrization of this flow in a way which
stresses the difference with the conventional stationary Roberts
flow.

The effective drift velocity of the magnetic field ${\bf V}^{\rm eff}$ is given by
\begin{eqnarray}
V^{\rm eff}_{x} &=& - {\tau \sigma \over 4l} \, \sin (2 x)  \;,
\quad
V^{\rm eff}_{y} = - {\tau \sigma \over 4l} \, \sin (2 y)  \;,
\nonumber\\
V^{\rm eff}_{z} &=& 0 \;,
\label{C4}
\end{eqnarray}
[see Eqs.~(\ref{R4}) and~(\ref{C3}) in Appendix B]. The effective drift velocity of the magnetic field determines the diamagnetic (or paramagnetic) effect. Using Eqs.~(\ref{R3}) and~(\ref{C3}) we determine the symmetric tensor $\eta_{ij}$ for the Roberts flow:
\begin{eqnarray}
\eta_{xx} &=& {\tau \sigma \over 2} \, \sin^2 y  \;,
\quad
\eta_{xy} = {\tau \sigma \over 8} \, \sin (2 x) \, \sin (2 y) \;,
\nonumber\\
\eta_{yy} &=& {\tau \sigma \over 2} \, \sin^2 x  \;,
\quad
\eta_{yz} = - {\tau \sigma \over 4} \, \sin^2 y \, \sin (2 x) \;,
\nonumber\\
\eta_{zz} &=& {\tau \sigma \over 4} \, \big[1 - \cos (2 x) \, \cos (2 y)] \;,
\nonumber\\
\eta_{xz} &=& {\tau \sigma \over 4} \, \sin^2 x \, \sin (2 y) \; .
\label{C5}
\end{eqnarray}
Equation~(\ref{R5}) yields the tensor ${\kappa}_{ijk} = - \tau \varepsilon_{ijm} u_{mk}$, where the second moment for the velocity field  $u_{ij} = \langle v_i v_j \rangle$ is determined by Eq.~(\ref{C3}).

The spatial averaging $\langle ... \rangle^{(sp)}$ of the electromotive force yields $\langle \bec{\cal E} \rangle^{(sp)} = \bec{\cal E}^{(1)} + \bec{\cal E} ^{(2)}$, where:
\begin{eqnarray}
\bec{\cal E}^{(1)} = {\tau \sigma \over 4} \, \big[ 2 (\delta_{ij} - e_i e_j) B_j \, l^{-1} - \bec{\nabla} {\bf \times} {\bf B} \big] \;,
\label{C6}
\end{eqnarray}
and the second contribution to the electromotive force $\bec{\cal E}^{(2)}$ is given by Eq.~(\ref{M5}) in Appendix B. This contribution causes decrease of the total turbulent magnetic diffusion and increase of the total $\alpha$ effect. However these effects are small [see Eqs.~(\ref{M6}) and~(\ref{M7})]. This implies that the equation for the mean magnetic field evolution for the random Roberts flow reads:
\begin{eqnarray}
{\partial B_i \over \partial t} = {\tau \sigma \over 2l} \, (\delta_{ij} - e_i e_j) \,
\big( \bec{\nabla} \times {\bf B} \big)_j + \Big(\eta + {\tau \sigma \over 4} \Big) \Delta B_i  \, .
\label{C7}
\end{eqnarray}

\section{Discussion}

A natural step in further analysis is to solve the mean-field
equations with transport coefficients obtained for the random $ABC$ flow and Roberts flow. Straightforward calculations in Fourier space
yield the growth rates of the large-scale magnetic field for the $ABC$ flow:
\begin{eqnarray}
\gamma = K  \biggl[{\tau \sigma_{_{A}} \over l} - (\tau \sigma_{_{A}}  +  \eta) K\biggr] \;,
 \label{D1}
 \end{eqnarray}
(see Eq. (\ref{B7})), and for the Roberts flow:

\begin{eqnarray}
\gamma = K  \biggl[{\tau {\sigma} \over 2l} - \Big({\tau \sigma \over 4} + \eta\Big) K\biggr] \;,
 \label{D2}
 \end{eqnarray}
(see Eq.~(\ref{C7})), where $K = 2 \pi / L_B$ is wave-number and $L_B$ is the characteristic scale of the magnetic field variations.
Equations~(\ref{D1}) and~(\ref{D2}) imply that the large-scale magnetic field grows in the same time scale for both flows. Moreover,
the growth rates obtained are similar to that for the mean-field dynamo in developed mirror asymmetric turbulence with $\alpha$-effect.

Therefore, our study confirms Zeldovich idea that the time dependence (e.g., in the form of random in time fluctuations)
remove the obstacle in large-scale dynamo action of classic stationary flows and provide dynamos. Moreover, we do not see
any necessity for an instantaneous flow geometry to be topologically complex. Of course, trajectories of the flow particles
remain chaotic due to random nature of governing parameters.

In this study we choose parametrization of the fluctuating $ABC$ and Roberts flows in a way to vanish the mean velocities and stress the
difference with conventional stationary $ABC$ and Roberts flows. Therefore, the dynamo effects isolated are associated with fluctuations only. When $\langle A \rangle \not= 0$ (and $\langle C \rangle \not= 0)$ the standard ${\bf U} {\bf \times} {\bf B}$ term in the mean-field dynamo equation vanishes, where ${\bf U} = \langle \langle {\bf v} \rangle \rangle^{(sp)}$ and ${\bf B} = \langle \langle {\bf H} \rangle \rangle^{(sp)}$  are the mean velocity and magnetic fields. On the other hand, there can be a non-vanishing contribution to the mean-field dynamo equation caused by the term $\langle \langle {\bf v} \rangle {\bf \times} \tilde {\bf b} \rangle^{(sp)}$, where $\tilde {\bf b}$ is determined by Eq.~(\ref{R6}) with an additional source term $\propto - \bec{\nabla} {\bf \times} [\langle {\bf v} \rangle {\bf \times} {\bf B}]$. However, an account for the effect of the mean fluid velocity on the mean-field dynamo is out of the scope of the present study.

Note that stationary deterministic flows (like the $ABC$-flow with constant coefficients) cannot cause dynamo of the large-scale magnetic field, i.e., stationary flows cannot generate magnetic field in the scales which are larger than the characteristic scales of the fluid flow field. On the other hand, numerical experiments \cite{AK83,GF86} show that the $ABC$-flow with constant coefficients excites a magnetic field in the scales which are smaller than the scales of the fluid flow field. The generation of the magnetic field by the stationary  $ABC$-flow requires the magnetic Reynolds numbers, ${\rm Rm}$, which are larger than 9. In particular, as follows from numerical experiments \cite{AK83,GF86} the magnetic field is excited by the $ABC$-flow within two ranges of the magnetic Reynolds numbers: the first one is for $9 < {\rm Rm} < 17.5$ and the second range is for ${\rm Rm} > 27$. Stationary Roberts flow cannot generate magnetic field. In contrast, in the present study we have shown that the generation of large-scale magnetic fields by random $ABC$ or Roberts flows can be possible even for small magnetic Reynolds numbers.

The analysis undertaken in the present study is not addressed to clarify the problems of nonlinear dynamo saturation and dynamics of small-scale magnetic fields associated with dynamo action (see, e.g., \cite{BS05,TC08}). We highly appreciate the importance of these problems, however this is out of the scope of this paper. Main goal of the present study is to demonstrate a possibility for a mean-field dynamo action (in the scales which are larger than the characteristic scales of the fluid flow field) in random ABC and Roberts flows in the framework of the kinematic approach.

Based on the Roberts flow in a finite cylindrical domain
\cite{RO70,B78}, one of the first successful laboratory dynamo
experiments was performed in Karlsruhe \cite{SM01,MS02,MSH04}. These
experiments demonstrate a generation of magnetic field in a
cylindrical container filled with liquid sodium in which by means of
guide tubes counter-rotating and counter-current spiral vortices are
established. The dynamo in the Karlsruhe-type experiment is
self-exciting and the magnetic field saturates at a mean value for
fixed supercritical flow rates \cite{SM01,MS02,MSH04}. The random
Roberts flow might be created in the Karlsruhe-type experimental set-up
when the pumping of energy in flow tubes has a random component. The
present study might be relevant to such dynamo experiment.

Solar super-granular flow structures \cite{P82} can be considered as ensemble of random cells. Their collective effect on generation of the solar magnetic field at scales which are larger than the sizes of super-granulations, is very important. It is plausible to suggest that the dynamo in random $ABC$ or Roberts flows can mimic the effect of random super-granulations on the large-scale solar dynamos.

\begin{acknowledgments}
DT and DS are grateful to RFBR for financial support under grant
07-02-00127.
\end{acknowledgments}

\appendix

\section{The electromotive force for random $ABC$-flow}

Using Eq.~(\ref{B1}) we get the tensor $u_{ij} = \langle v_i v_j \rangle$ for random $ABC$-flow:
\begin{eqnarray}
u_{xx} &=& \sigma_{_{B}} \, \cos^2 y + \sigma_{_{C}} \, \sin^2 z \;,
\quad
u_{xy} = {\sigma_{_{C}} \over 2} \, \sin (2 z) \;,
\nonumber\\
u_{yy} &=& \sigma_{_{C}} \, \cos^2 z + \sigma_{_{A}} \, \sin^2 x  \;,
\quad
u_{yz} = {\sigma_{_{A}} \over 2} \, \sin (2 x) \;,
\nonumber\\
u_{zz} &=& \sigma_{_{A}} \, \cos^2 x + \sigma_{_{B}} \, \sin^2 y  \;,
\quad
u_{xz} = {\sigma_{_{B}} \over 2} \, \sin (2 y) \;,
\nonumber\\
\label{B3}
\langle {\bf v}^2 \rangle &=& \sigma_{_{A}} + \sigma_{_{B}} + \sigma_{_{C}} \; .
\end{eqnarray}

Now we consider for simplicity the case $\sigma_{_{A}} = \sigma_{_{B}}
= \sigma_{_{C}}$. Equation~(\ref{R6}) yields the equation for the magnetic field $\tilde {\bf b}$ for the random $ABC$-flow that is in Fourier space reads:
\begin{eqnarray}
D_{ij} \tilde b_j = - {l^2 \over (\eta + \tau \sigma_{_{A}}) k^2} \, I_i \;,
\label{L1}
\end{eqnarray}
where
\begin{eqnarray}
D_{ij} = \delta_{ij} - {i \over k^2} \varepsilon_{ijm} k_m \;,
\label{L2}
\end{eqnarray}
${\bf k}$ is the dimensionless wave vector measured in the units of $l^{-1}$ and $l$ is the characteristic scale of the velocity field variations, $I_i$ is determined by Eq.~(\ref{R7}), and we have taken into account Eqs.~(\ref{B6}) and~(\ref{B7}). The solution of Eq.~(\ref{L1}) reads:
\begin{eqnarray}
\tilde b_i = - {l^2 \over (\eta + \tau \sigma_{_{A}}) k^2} \, D_{ij}^{-1} I_j \;,
\label{L3}
\end{eqnarray}
where the inverse operator $D_{ij}^{-1}$ is given by
\begin{eqnarray}
D_{ij}^{-1} = (k^2 - 1)^{-1} \Big(k^2 \delta_{ij} + i \varepsilon_{ijm} k_m - {k_i k_j \over k^2} \Big) \; .
\nonumber\\
\label{L4}
\end{eqnarray}

Straightforward calculations using Eqs.~(\ref{L3}) and (\ref{L4}) yield the second moments $\langle \tilde {\eta}_{ij} (\bec{\nabla} {\bf \times} \tilde {\bf b})_{j} \rangle^{(sp)}$ and
$\langle \tilde {\kappa}_{ijk} ({\partial \tilde b})_{jk} \rangle^{(sp)}$ for random $ABC$ flow:
\begin{eqnarray}
\langle \tilde {\eta}_{ij} (\bec{\nabla} {\bf \times} \tilde {\bf b})_{j} \rangle^{(sp)} &=& -
{\eta^{\rm eff}_{_{A}} \over 6} (\bec{\nabla} {\bf \times} {\bf B})_{i} \;,
\label{L5}\\
\langle \tilde {\kappa}_{ijk} ({\partial \tilde b})_{jk} \rangle^{(sp)} &=& {\eta^{\rm eff}_{_{A}}
\over 6} (\bec{\nabla} {\bf \times} {\bf B})_{i} \;,
\label{L6}
\end{eqnarray}
where $\eta^{\rm eff}_{_{A}}=\tau^2 \sigma_{_{A}}^2 / (\eta + \tau \sigma_{_{A}})$, and we have taken into account that $k_x = k_y = 2$. Note also that for the random $ABC$-flow $\langle \tilde {\alpha}_{ij} \tilde b_{j} \rangle^{(sp)} = 0$ and $\langle \tilde {\bf V}^{\rm eff} {\bf \times} \tilde {\bf b} \rangle^{(sp)}  = 0$, because $\tilde {\alpha}_{ij} = 0$ and $\tilde {\bf V}^{\rm eff} = 0$.

\section{The electromotive force for random Roberts flow}

Using Eq.~(\ref{R3}) we get the tensor $u_{ij}= \langle v_i v_j \rangle$ for random Roberts flow:
\begin{eqnarray}
u_{xx} &=& \sigma \, \sin^2 x \, \cos^2 y \;,
\quad
u_{xy} = - {\sigma \over 4} \, \sin (2 x) \, \sin (2 y) \;,
\nonumber\\
u_{yy} &=& \sigma \, \cos^2 x \, \sin^2 y \;,
\quad
u_{yz} = {\sigma \over 2} \, \sin^2 y \, \sin (2 x) \;,
\nonumber\\
u_{zz} &=& \sigma \, \sin^2 x \, \sin^2 y  \;,
\quad
u_{xz} = - {\sigma \over 2} \, \sin^2 x \, \sin (2 y) \;,
\nonumber\\
\langle {\bf v}^2 \rangle &=& \sigma \big(\sin^2 x + \cos^2 x \, \sin^2 y \big) \; .
\label{C3}
\end{eqnarray}

The solution of Eq.~(\ref{R6}) yields the magnetic field $\tilde {\bf b}$ for random Roberts flow that is determined by the following equation in Fourier space:
\begin{eqnarray}
\tilde b_i &=& {l^2 \over k^2} \, \Big(\eta + {\tau \sigma \over 4} \Big)^{-1} \Big\{I_i + {2 \over k^2} \Big[({\bf e} {\bf \cdot} \bec{\nabla}) \,
\big({\bf e} {\bf \times} {\bf I})_i
\nonumber\\
&&- \big({\bf e} {\bf \times} \bec{\nabla}\big)_i \, ({\bf e} {\bf \cdot} {\bf I}) + {2 \over k^2} ({\bf e} {\bf \cdot} \bec{\nabla}) \, \nabla_i \, ({\bf e} {\bf \cdot} {\bf I}) \Big] \Big\}  \;,
\label{M1}
\end{eqnarray}
where $I_i$ is given by Eq.~(\ref{R7}), ${\bf k}$ is the dimensionless wave vector measured in the units of $l^{-1}$. Here we have taken into account Eqs.~(\ref{C6}) and~(\ref{C7}). Straightforward calculations yield
\begin{eqnarray}
\langle \tilde {\alpha}_{xj} \tilde b_{j} \rangle^{(sp)} &=&
{\eta^{\rm eff} \over 32} {\nabla}_y B_{z} \;,
\quad
\langle \tilde {\alpha}_{yj} \tilde b_{j} \rangle^{(sp)} =
{\eta^{\rm eff} \over 32} {\nabla}_x B_{z},
\nonumber\\
\langle \tilde {\alpha}_{zj} \tilde b_{j} \rangle^{(sp)} &=& {\eta^{\rm eff} \over 16} \big[2(\bec{\nabla} {\bf \times} {\bf B})_{z} + {\nabla}_y B_{x} \big] \;,
\label{MM2}
\end{eqnarray}
\begin{eqnarray}
\langle (\tilde {\bf V}^{\rm eff} {\bf \times} \tilde {\bf b})_x \rangle^{(sp)} &=& - {\eta^{\rm eff} \over 32}
{\nabla}_y B_{z} \;,
\nonumber\\
\langle (\tilde {\bf V}^{\rm eff} {\bf \times} \tilde {\bf b})_y \rangle^{(sp)} &=& {\eta^{\rm eff} \over 32}
{\nabla}_x B_{z} \;,
\nonumber\\
\langle (\tilde {\bf V}^{\rm eff} {\bf \times} \tilde {\bf b})_z \rangle^{(sp)} &=& {\eta^{\rm eff} \over 32} (\bec{\nabla} {\bf \times} {\bf B})_{z} \;,
\label{M2}
\end{eqnarray}
\begin{eqnarray}
\langle \tilde {\eta}_{xj} (\bec{\nabla} {\bf \times} \tilde {\bf b})_{j} \rangle^{(sp)} &=& -
{\eta^{\rm eff} \over 64} [6(\bec{\nabla} {\bf \times} {\bf B})_{x} + 5 {\nabla}_z B_{y}] \;,
\nonumber\\
\langle \tilde {\eta}_{yj} (\bec{\nabla} {\bf \times} \tilde {\bf b})_{j} \rangle^{(sp)} &=&
{\eta^{\rm eff} \over 64} [(\bec{\nabla} {\bf \times} {\bf B})_{y} - 5 {\nabla}_x B_{z}] \;,
\nonumber\\
\langle \tilde {\eta}_{zj} (\bec{\nabla} {\bf \times} \tilde {\bf b})_{j} \rangle^{(sp)} &=&
{\eta^{\rm eff} \over 64} [(\bec{\nabla} {\bf \times} {\bf B})_{z} +4 \nabla_y B_x
\nonumber\\
&&- 8 \, l^{-1} B_{z}] \;,
\label{M3}
\end{eqnarray}
and
\begin{eqnarray}
\langle \tilde {\kappa}_{xjk} ({\partial \tilde b})_{jk} \rangle^{(sp)} &=& - {\eta^{\rm eff} \over 64} [3 (\bec{\nabla} {\bf \times} {\bf B})_{x} + 6\nabla_z B_z + \nabla_x B_x
\nonumber\\
&&+4 \, l^{-1} B_{x}] \;,
\nonumber\\
\langle \tilde {\kappa}_{yjk} ({\partial \tilde b})_{jk} \rangle^{(sp)} &=& - {\eta^{\rm eff} \over 64} [3 (\bec{\nabla} {\bf \times} {\bf B})_{y} + 6\nabla_x B_x + \nabla_y B_y
\nonumber\\
&&+4 \, l^{-1} B_{y}] \;,
\nonumber\\
\langle \tilde {\kappa}_{zjk} ({\partial \tilde b})_{jk} \rangle^{(sp)} &=& - {\eta^{\rm eff} \over 64} [13 (\bec{\nabla} {\bf \times} {\bf B})_{z} - 2\nabla_y B_x
\nonumber\\
&&- 8 \, l^{-1} B_{z}] \;,
\label{M4}
\end{eqnarray}
where $\eta^{\rm eff}=\tau^2 \sigma^2 / (4\eta + \tau \sigma)$, and we have taken into account that $k_x = k_y = 2$. The contribution to the electromotive force, $\bec{\cal E} ^{(2)}$, caused by the second moments $\langle \tilde {\alpha}_{ij} \tilde b_{j} \rangle^{(sp)}$, $\, \langle \tilde {\eta}_{ij} (\bec{\nabla} {\bf \times} \tilde {\bf b})_{j} \rangle^{(sp)}$, is given by
\begin{eqnarray}
{\cal E}^{(2)}_x &=& {\eta^{\rm eff} \over 64} [9 (\bec{\nabla} {\bf \times} {\bf B})_{x} +5 \nabla_z B_y + 6\nabla_z B_z + \nabla_x B_x
\nonumber\\
&& +4 \, l^{-1} B_{x}] \;,
\nonumber\\
{\cal E}^{(2)}_y &=& {\eta^{\rm eff} \over 64} [2 (\bec{\nabla} {\bf \times} {\bf B})_{y} + 9 \nabla_x B_z + 6\nabla_x B_x + \nabla_y B_y
\nonumber\\
&& +4 \, l^{-1} B_{y}] \;,
\nonumber\\
{\cal E}^{(2)}_z &=& {\eta^{\rm eff} \over 32} [11 (\bec{\nabla} {\bf \times} {\bf B})_{z} - \nabla_y B_x] \; .
\label{M5}
\end{eqnarray}
This implies that the diagonal components of the total turbulent magnetic diffusion tensor are
\begin{eqnarray}
\eta_{xx}^{\rm tot} &=& {\tau \sigma \over 4} \, \Big[1 - {9 \over 64} {\rm Rm}\Big] \;,
\nonumber\\
\eta_{yy}^{\rm tot} &=& {\tau \sigma \over 4} \, \Big[1 - {1 \over 32} {\rm Rm}\Big] \;,
\nonumber\\
\eta_{zz}^{\rm tot} &=& {\tau \sigma \over 4} \, \Big[1 - {11 \over 32} {\rm Rm}\Big] \;,
\label{M6}
\end{eqnarray}
while the diagonal components of the total $\alpha$ tensor are given by
\begin{eqnarray}
\alpha_{xx}^{\rm tot} = \alpha_{yy}^{\rm tot} = {\tau \sigma \over 2l} \, \Big[1 + {1 \over 32} {\rm Rm}\Big] \;,
\label{M7}
\end{eqnarray}
and $\alpha_{zz}^{\rm tot} =0$. Here ${\rm Rm}=\tau \sigma / \eta \ll 1$ and we took into account Eqs.~(\ref{C6}) and~(\ref{M5}).

\end{document}